\documentclass[mathleft
]{an}
\usepackage{graphicx}
\usepackage{times}
\begin{document}

\Pagespan{789}{}
\Yearpublication{2010}%
\Yearsubmission{2010}%
\Month{11}%
\Volume{999}%
\Issue{88}%
 \DOI{10.1002/asna.201011374}%

\title{Center to limb variation of penumbral Stokes V profiles.}

\author{M. Franz\inst{1}\fnmsep\thanks{Corresponding author:
  \email{morten@kis.uni-freiburg.de}\newline}
\and  R. Schlichenmaier\inst{1}
}
\titlerunning{Center to limb variation of penumbral Stokes V profiles}
\authorrunning{M. Franz \and R. Schlichenmaier}
\institute{
Kiepenheuer Institut f\"ur Sonnenphysik,
Sch\"oneckstra\ss{}e 6, D-79104 Freiburg\\}

\received{19 Jan 2010}
\accepted{29 Mar 2010}
\publonline{}
\sloppy


\abstract{We investigated the horizontal and the vertical component of the Evershed flow (EF). To this end, we computed average Stokes V profiles for various velocity classes in penumbrae at different heliocentric angles. Our results show that for blueshifted profiles an additional lobe with the same polarity as the spot is present in the blue side of the average Stokes V profile. The amplitude of the additional lobe grows with increasing blueshift and with increasing heliocentric angle. For small redshifts, the profiles show an additional lobe with the opposite polarity as the spot on the red side of the average Stokes V profile. Even at disk center, the original polarity of the average Stokes V profile is reversed for strong redshifts. The transition between the different types of Stokes V profiles is continuous and indicates that not only the vertical, but also the horizontal EF is a magnetized stream of plasma in a magnetic background field.}

\maketitle

\section{Introduction}
The Evershed flow (EF) is known from the shift of penumbral photospheric lines that increases with the heliocentric angle (\cite{Evershed1909}). The strength of the shift of photospheric lines is decreasing with the strength of the line (\cite{StJohns1913}).
Observations with higher spectral resolution show that penumbral lines are asymmetric insofar as the line wing is stronger shifted than the line core (\cite{Bumba1960a}, \cite{Bumba1960b}, \cite{Servajean1961}). Sometimes, the line asymmetry is so large that additional line 'satellites' appear (\cite{Wiehr1995}). Considering these results, the EF has been interpreted as a depth-dependent radial outflow of plasma in the penumbra: The flow is present in the deep photosphere, thereby causing the asymmetry of the line wing or the appearance of line 'satellites'  (\cite{Maltby1964}, see also \cite{schlichenmaier+etal2004}).

However, these observations involve non-magnetic lines or do not consider all Stokes parameters. Spectropolarimetric observations are important though, since they contain information about the magnetic field and the gradients of the atmospheric parameters along the line of sight (LOS). One example are the so called 'crossover' profiles - i.e. Stokes V profiles with additional lobes in the blue or red wing of a regular Stokes V profile - found along the penumbral magnetic neutral line (MNL)
(\cite{SanchezAlmeida1992}, \cite{Schlichenmaier2002}). Using HINODE data, it is possible to identify 'crossover' profiles not only along the MNL, but throughout the penumbra (\cite{SanchezAlmeida2009}).
These profiles can be explained assuming a two component model, in which the superposition of the signals from the two components will produce asymmetric Stokes I and 'crossover' Stokes V profiles. Even though such profiles are known for some time (\cite{Katz1972}), a comprehensive investigation of these profiles in the sunspot penumbrae, especially their dependence on Doppler velocity and heliocentric angle, has yet to be conducted.

\section{Observation and data calibration}
The sunspot investigated in this study was observed at disk center ($\Theta = 3^{\circ}$) on the 5$^{\rm{th}}$ of January 2007 and close to the limb ($\Theta = 47^{\circ}$) on the 9$^{\rm{th}}$ of January 2007. The data was obtained with the spectropolarimeter (\cite{Lites2001}) of the solar optical telescope (\cite{Tsuneta2008}) onboard {\it{{Hinode}}}. The spatial resolution of the images is 0."32, the spectral sampling is 2.15 pm/pixel and the 1$\sigma$ noise level of the spectra is of the order of $10^{-3}$.

For wavelength calibration, we assumed that no plasma motion is present in the line forming region above the umbra, if the corresponding Stokes V profile is completely antisymmetric (\cite{Auer1978}). Therefore, the center position of antisymmetric Stokes V profiles can be used to define a rest frame on the solar surface (\cite{Rezaei2006}).

To create maps of Doppler velocity, a bisector between 70 \% and 90 \% line intensity was computed from Fe 630.15 mn, and its average position was compared to the center position of antisymmetric Stokes V profiles of the umbra. Note that the resulting velocities refer to an average gas velocity of the layer where the line wings are forming,§ § and therefore underestimate the actual flow velocity. For a more detailed description of the calibration process as well as the computation of the Doppler maps including an error discussion please refer to our previous paper (\cite{Franz2009}).

\section{Results}

\begin{figure*}[htb]
	\begin{center}
		\includegraphics[width=0.999\textwidth]{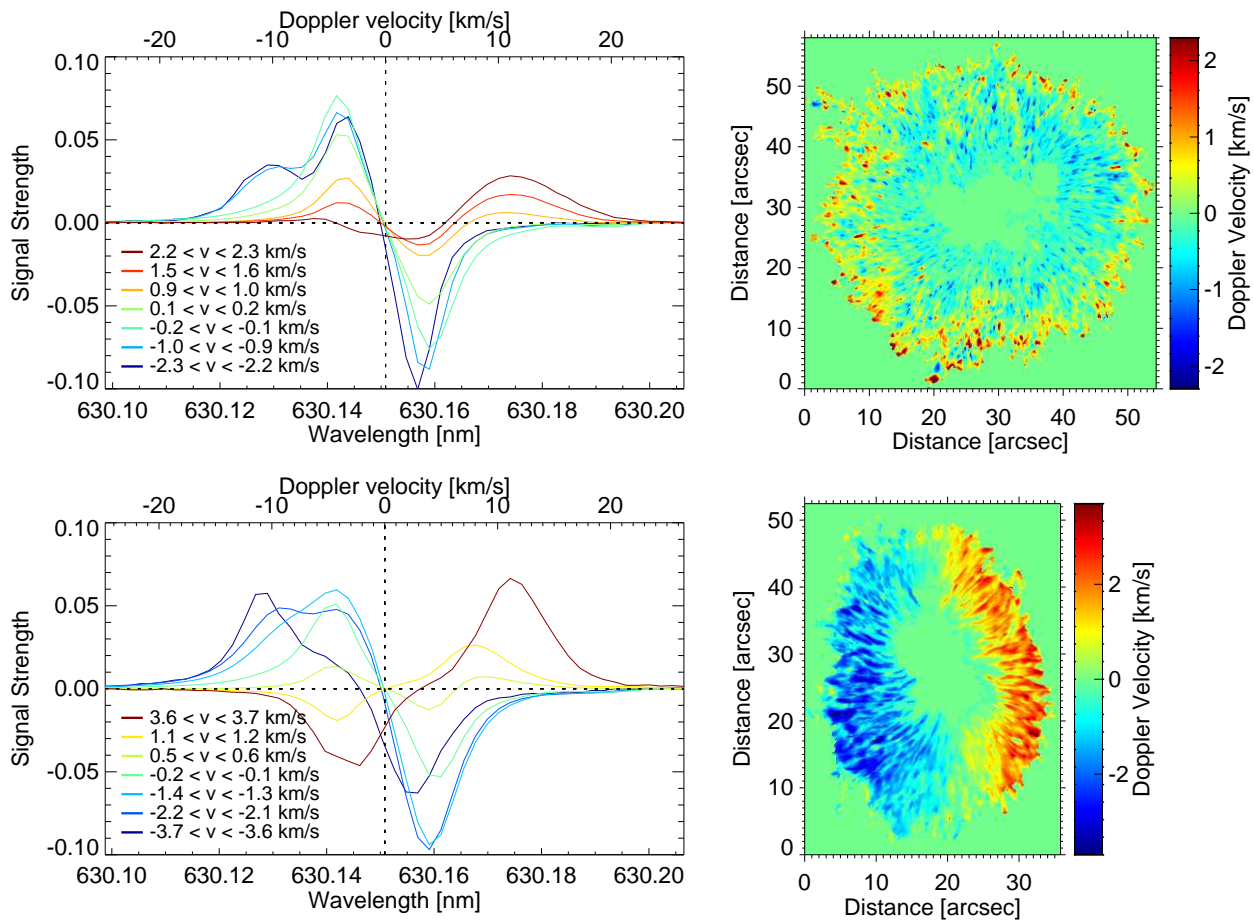}
		\caption{Figures in the upper row refer to the sunspot at $\Theta = 3^{\circ}$, results in the lower row refer to the same sunspot at $\Theta = 47^{\circ}$. The left column shows the averaged Stokes V profiles for selected velocity classes, with colors indicating the strength of red- or blueshift. The images in the right column depict the Doppler velocities in the penumbra at both heliocentric angles.}
		\label{Franz_fig00}
	\end{center}
\end{figure*}

The upper right panel of Fig.~\ref{Franz_fig00} shows the Doppler velocities in the penumbra at disk center ($\Theta = 3^{\circ}$). The Doppler shift in this map shows an up- or downflow and has been cropped at a corresponding velocity of ${\rm{v}} = \pm 2.3$ km/s. The upper left panel shows seven averaged penumbral Stokes V profiles: One of these profiles represents the average of all Stokes V profiles in a certain Doppler velocity class. Different colors indicate different velocity classes, e.g. the blue profile is the average of all profiles with an upflow $-0.9$ {\rm{km/s}} $ \ge {\rm{v}} \ge -1.0$ {\rm{km/s}}. Several features are evident from this plot:

If the blueshift, i.e. the upflow velocity, is small, the Stokes V profile appears antisymmetric (dark green). As the upflow velocity increases, a shoulder becomes apparent on the blue side of the profile, while the amplitude of the blue wing decreases (blue). For very strong upflows (dark blue), this shoulder transforms into a second lobe with the same polarity as the main component. Profiles representing downflows behave differently. As the downflow velocity increases, the amplitude of the Stokes V profiles decreases. Aditionally, a second lobe with opposite polarity appears on the red side of the profile (yellow). For large downflow velocities (${\rm{v}} > 2.2$ km/s), the polarity of the Stokes V profile is reversed (dark red).

The cropped Doppler velocity map (${\rm{v}} = \pm 3.7$ km/s) in the lower right panel of Fig.~\ref{Franz_fig00} shows a penumbra off disk center ($\Theta = 47^{\circ}$). At this heliocentric angle not the up- or downflow, but the horizontal component of the EF is the dominant contribution to the Doppler shift. The average penumbral Stokes V profiles for different velocity classes are depicted in the lower left panel. Their behavior is similar to those at disk center:

If the Doppler shift is small, the Stokes V profiles appear antisymmetric (turquoise). For increasing blueshift, a shoulder of the same polarity is seen on the blue side of the profile (light blue). The amplitude of the second component is stronger as at disk center (blue), and for ${\rm{v}} < -2$ km/s its amplitude becomes the dominant one (dark blue). The main difference to redshifted penumbral Stokes V profiles at disk center is that an opposite polarity lobe is already present at ${\rm{v}} > 0.5$ km/s (green) and that for all ${\rm{v}} > 1.1$ km/s, the polarity of Stokes V is completely reversed (yellow). 

It should be mentioned that the averaging process hides certain features. While the 'crossover' effect becomes apparent in the averaged profiles only at larger velocities, some Stokes V profiles already show this effect at $0.1 > {\rm{v}} > 0.2$ km/s. However, the transition between the Stokes V profiles of different velocity classes occurs continuously and is not an artifact of the averaging process.

\section{Discussion and conclusion}

Following the concept of \cite{Maltby1964}, we interpret our results in terms of a simple atmospheric model. To explain any area asymmetries in penumbral Stokes V profiles, it is necessary to assume gradients of the atmospheric parameters within the line forming region (\cite{Landi1983}). In this scenario, the gradients may be described by assuming a stack of two atmospheric layers, i.e. two overlaying components with different atmospheric parameters. Many properties of the Stokes V profiles in Fig.~\ref{Franz_fig00} can be understood in terms of this two component atmospheric model, which shall be discussed in the following. Besides a difference in velocity, a gradient in any other atmospheric parameters - e.g. the magnetic field strength - will influence the resulting Stokes profiles. One may compute synthetic profiles or use inversion codes to model these gradients. However, this procedure is beyond the scope of this investigation. Therefore, we focus on the difference in velocity, which is essential to produce line 'satellites'.

{\bf{Disk center ($\Theta = 3^{\circ}$):}} Let one atmospheric component be magnetized plasma at rest and let the other one represent magnetized plasma in motion. A superposition of these two components will result in a two lobed Stokes V profile with an area asymmetry if the differences in the velocity ($\Delta$v) of the two components is small. If $\Delta$v is large and if the magnetic field in both components has the same polarity, an additional lobe with the same polarity will appear in the resulting Stokes V profile. It is located on the blue side of the blue lobe, if the plasma moves towards the observer, and it is on the red side of the red lobe for a plasma moving away from the observer. A large $\Delta$v and a magnetic field of opposite polarity in the two layers will result in an additional lobe of opposite polarity in the StokesV profile. 

The change of amplitude for different velocity classes can be understood, if the spatial position in the penumbra is taken into account. At disk center, redshifted profiles are present in the downflow regions in the outer penumbra, while blueshifted profiles are located in the inner penumbra. Since the inclination of the magnetic field increases with radial distance form the umbra whereas the strength of the magnetic field decreases at the same time (e.g., \cite{Keppens1996}), regular Stokes V profiles from upflow regions are expected to have a larger amplitude than profiles from downflow regions.

{\bf{Large heliocentric angles ($\Theta = 47^{\circ}$):}} Away from disk center, geometric effects, i.e. the shape of the global magnetic field in the penumbra, have to be considered. Since part of the magnetic field of the limb-side penumbra appears orthogonal to the LOS, only the transverse magnetic field shows a non-zero fieldstrength. This results in the MNL, across which the Stokes V profile changes its polarity. As the velocity of the EF peaks in the outer penumbra (\cite{BellotRubio2003}, \cite{tritschler+al2004}) -- in this case beyond the MNL --, it is evident that strongly redshifted Stokes V profiles show the opposite polarity, while weakly redshifted profiles - inner limb-side-penumbra - show the same polarity as the spot.

However, the interpretation of the amplitudes of blue\-shifted profiles from the center-side penumbra faces difficulties. If we focus on the blue lobe of 'crossover' Stokes V profiles, we find for all profiles with $\rm{v} < -2$ km/s, an amplitude of the (blue-shifted) flow component which is larger than the amplitude of the component at rest. This amplitude ratio is just opposite to the one found in profiles from penumbrae disk center. Since the inclination between the LOS and the magnetic field of the streaming component is found to be larger than the inclination between the LOS and the magnetic field of the component at rest (\cite{BellotRubio2003}, \cite{schmidt+schl2000}), the Stokes V amplitude of the streaming component should be smaller than that of the component at rest. At the moment we do not have a final explanation for this behavior. We can only speculate that it might be due to projection effects: The large heliocentric angle leads to a smaller inclination between the LOS and the flow component. Therefore, the contribution of this flow is larger when compared to penumbrae at disk center. This might explain the observed behavior of the Stokes V amplitudes.

{\bf{Conclusion:}} If we interpret the averaged Stokes V profile for the disk center penumbra in terms of a simple two component model, it is stringent to assume that up- and downflows in the penumbra are magnetized. Otherwise, it is impossible to explain the additional lobes present in the averaged Stokes V profiles from the penumbra at disk center. Consequently, the horizontal component of the EF, dominating the velocity signal off disk center, needs to be magnetized as well (cf. lower left panel in Fig. \ref{Franz_fig00}).

\begin{acknowledgements}
We thank W. Schmidt for his valuable comments on the manuscript. Part of this work was supported by the \emph{Deut\-sche For\-schungs\-ge\-mein\-schaft}, DFG project no. Schl.~514/\mbox{3-1.} {\it{Hinode}} is a Japanese mission developed and launched by ISAS/ JAXA, with NAOJ as domestic partner and NASA and STFC (UK) as international partners. It is operated by these agencies in cooperation with ESA and NSC (Norway).
\end{acknowledgements}


\end{document}